\documentclass[aps,prl,amsmath,amssymb,floatfix,twocolumn]{revtex4-1}
\usepackage{bbold}
\usepackage{graphicx}
\usepackage{subfigure}
\usepackage{color}
\usepackage{hyperref}

\begin{document}

\title{Effective field theory, chiral anomaly and vortex zero modes for odd parity topological superconducting state of three dimensional Dirac materials}

\author{Pallab Goswami}
\affiliation{National High Magnetic Field Laboratory and Florida State University, Tallahassee, Florida 32310, USA}
\author{Bitan Roy}
\affiliation{National High Magnetic Field Laboratory and Florida State University, Tallahassee, Florida 32310, USA}

\begin{abstract}
The low energy quasiparticle dispersion of various narrow gap and gapless semiconductors are respectively described by three dimensional massive and massless Dirac fermions. The three dimensional Dirac spinor structure admits a time-reversal invariant, odd parity and Lorentz pseudoscalar topological superconducting state. Here we derive the effective field theory of this topological paired state for massless Dirac fermions in the presence of a fluctuating Zeeman term, which appears as a chiral gauge field. The effective theory consists of a mixed electromagnetic and chiral anomaly term in the bulk, and a combination of pure and mixed anomalies for the surface. In this paper we demonstrate the existence of fermion zero modes in the dilute vortex phase under generic conditions. Guided by the existence of the zero modes and its intimate connection with the anomaly, we propose an effective topological field theory in the presence of Dirac mass. We briefly discuss the experimental consequences of the effective field theory and the zero modes for the low temperature unconventional superconducting states of $\mathrm{Cu}_x\mathrm{Bi}_2\mathrm{Se}_3$ and $\mathrm{Sn}_{1-x}\mathrm{In}_x\mathrm{Te}$.

\end{abstract}

\maketitle

\emph{Introduction:} Recently there has been a surge of theoretical and experimental interest in the time-reversal invariant topological states of matter in three spatial dimensions\cite{Review1, Review2, Review3}. In contrast to the abundance of materials in which a three dimensional $Z_2$ strong topological insulator phase has been realized, there are only a few candidates for three dimensional, time-reversal invariant topological superfluid (TSF) and superconducting (TSC) states. Apart from the $^3He-B$ phase, which is a charge neutral TSF \cite{Review3, Volovik}, there is only one charged, odd parity TSC state, which has been proposed in Ref.~\onlinecite{FuBerg} for $\mathrm{Cu}_x\mathrm{Bi}_2\mathrm{Se}_3$ \cite{ObWray1, ObKriener, ObWray2}. The odd parity TSC state may also occur in other narrow gap semi-conductors and semi-metals with a Dirac like quasiparticle dispersion in the normal state e.g, $\mathrm{Sn}_{1-x}\mathrm{In}_x\mathrm{Te}$ \cite{ObSasaki}. Both of these paired states belong to class DIII in the periodic table for topological insulators and superconductors\cite{Altland, TenfoldRyu}. The BdG Hamiltonians for uniform topological paired states in class DIII possess a nontrivial $Z$ invariant ($N$)\cite{Review3, TenfoldRyu, FuBerg}. At an interface between two bulk states with different $N$, the jump $\Delta N$ governs the number of two component, two dimensional, massless Majorana fermions at the surface. The vortex physics of topological paired states is also supposed to be rich, due to the presence of gapless one dimensional modes along the vortex cores \cite{Review3, Volovik}.

The recent point contact spectroscopy measurements on $\mathrm{Cu}_x\mathrm{Bi}_2\mathrm{Se}_3$\cite{PCSasaki, PCKirzhner, PCChen} and $\mathrm{Sn}_{1-x}\mathrm{In}_x\mathrm{Te}$\cite{ObSasaki}, have revealed a zero bias conductance peak (ZBCP), which is consistent with the existence of gapless Majorana surface states. However surface Majorana fermions may also emerge for other possible odd parity superconducting states, which are gapless in the bulk (such states are not described by the $Z$ invariant of class DIII) \cite{PCSasaki}. The magnetization measurements on $\mathrm{Cu}_x\mathrm{Bi}_2\mathrm{Se}_3$ also point towards an unconventional paired state \cite{MagDas, MagBay, MagKriener}. On the other hand absence of ZBCP in a recent tunneling spectroscopy measurement has been interpreted in terms of a fully gapped, topologically trivial paired state \cite{ObLevy}. Therefore more experiments are needed to identify the pairing symmetry.

So far, most of the theoretical studies have focused on the nature of the surface states in the presence of bulk topological pairing \cite{ThSHao, ThSYamakage, ThSMichaeli, ThSHsieh, ThSChung}. The thermal Hall effect and its connection to gravitational chiral anomaly have been proposed as the appropriate topological response for the detection of the surface Majorana fermions (since charge is not a conserved quantity) in Refs.~\onlinecite{GravQi, GravRyu1, GravRyu2}. On the other hand there are only a few theoretical studies \cite{QiWitten, BFHanson} of the bulk electrodynamic properties, which can also provide valuable insight into the nature of the underlying paired state. In Ref.~\onlinecite{BFHanson} a BF theory augmented by the fermion zero modes have been proposed. In Ref.~\onlinecite{QiWitten} employing a higher dimensional (4+1)-d regularization, an axion electrodynamics action have been found as the effective field theory for (3+1)-d TSC state. However the importance of Zeeman coupling and the interplay between charge and spin currents have not been considered in these papers. In this work we first consider the simpler case of massless Dirac fermions, and treat the orbital and Zeeman coupling on the same footing. By employing Fujikawa's chiral rotation technique \cite{Fujikawa} we derive the effective field theory for the odd parity TSC state, which shows the presence of a mixed electromagnetic and chiral anomaly in the bulk, which becomes operative only in the vortex phase. We also find a combination of axial anomaly (pure electromagnetic + pure chiral anomalies) and mixed anomaly for the surface states. 

\emph{Normal state Hamiltonian:} The low energy, long wavelength quasi-particle spectrum of several narrow gap and gapless semiconductors are respectively described by four component massive and massless Dirac equations in three spatial dimensions. In the presence of electromagnetic field, the text-book Dirac Hamiltonian is
\begin{equation}
H_N=\int d^3x \bar{\psi}(-i\gamma^j \partial_j+e \gamma^{\mu}\mathcal{V}_\mu + m ) \psi,
\label{eq:1}
\end{equation}
where Fermi velocity $v$ has been set to unity, $m$ is the Dirac mass (band gap), $e$ is the electron's  charge, and $\mathcal{V}_{\mu}$ is the electromagnetic vector potential. The anticommuting $\gamma$ matrices satisfy $\{\gamma^{\mu},\gamma^{\nu}\}=2 g^{\mu \nu}$, where $g^{\mu \nu}=(1,-1,-1,-1)$ is the metric tensor and $\bar{\psi}=\psi^{\dagger}\gamma^0$.  Throughout the paper Greek and Latin indices will respectively correspond to  space-time and spatial components of a vector. For simplicity we are ignoring the spatial anisotropy (direction dependent Fermi velocities) inherited from the crystal symmetry. If we choose the four component spinor $\psi^{T}=(c^{+}_{\uparrow}, c^{+}_{\downarrow}, c^{-}_{\uparrow},c^{-}_{\downarrow})$, where $c^{\pm}_{s}$ respectively correspond to the annihilation operators for parity even and odd states, with spin projections $s= \uparrow, \downarrow $, the gamma matrices follow the so-called Dirac representation: $\gamma^0=\eta^3 \otimes \sigma^0$, $\gamma^j=i\eta^2 \otimes \sigma^j$, and $\gamma^5=i\gamma^0 \gamma^1 \gamma^2 \gamma^3=\eta^1 \otimes \sigma^0$, where $\{\gamma^5, \gamma^{\mu}\}=0$. The Hamiltonian $H_N$ is invariant with respect to the parity ($\mathcal{P}$), the time reversal ($\mathcal{T}$) and the charge conjugation ($\mathcal{C}$) transformations, respectively defined by $\mathcal{P} \psi (t,\mathbf{x})\mathcal{P}^{-1}=\gamma_0 \psi(t,-\mathbf{x})$, $\mathcal{T} \psi(t,\mathbf{x}) \mathcal{T}^{-1}=-\gamma_1\gamma_3 \psi(-t,\mathbf{x})$, and $\mathcal{C} \psi(x) \mathcal{C}^{-1}=-i\gamma_2 \psi^{\ast}(x)$. In the absence of gauge field, the massless Dirac Hamiltonian is invariant under the global chiral transformation $\psi \to e^{i(\theta/2)\gamma_5}\psi$, $\bar{\psi}\to \bar{\psi}e^{i(\theta/2) \gamma_5}$. This symmetry is broken by the Dirac mass at the classical level, and by the gauge field through quantum mechanical effects known as chiral anomaly.

Unlike the true relativistic field theory of electrons and positrons, the above Dirac description of the quasi-particles is just an emergent low energy theory. For electrons and positrons, the electromagnetic gauge field only appears in the Hamiltonian through covariant derivative, and this is also tied to the fact that gyromagnetic ratio for such fundamental particles is 2. For quasiparticles of a semi-conductor this is no longer true, and for this reason we need to consider the Zeeman coupling separately. When the high energy bands are integrated out in the presence of the electromagnetic field, the $g$ factors (actually a tensor) of two bands can be different \cite{LiuZeeman}. For this reason the Zeeman coupling (again ignoring the spatial anisotropy) can be described by
\begin{equation}
H_Z=\int d^3x \bar{\psi} (g_+ \gamma^j \gamma^5 B_j + g_- \epsilon^{ijk} \sigma_{ij} B_k) \psi,
\label{eq:2}
 \end{equation}
where $g_{+} \pm g_{-}$ respectively describe the $g$ factors of the parity even and odd bands, and $\mathbf{B}=\nabla \times \mathbf{V}$ is the magnetic field strength. In the above equation we have absorbed the Bohr magneton $\mu_B$ in the definition of $g_{\pm}$. Notice that \emph{the Zeeman term proportional to $g_{+}$ appears as a vector potential for the chiral gauge field}. We have introduced the matrices $\sigma^{jl}=\{\gamma^j, \gamma^l \}/2i$, and the $g_-$ term breaks chiral symmetry and charge conjugation symmetry. When the electromagnetic and the chiral gauge fields are simultaneously present, the chiral anomaly leads to the non-conservation of both chiral and electromagnetic currents (respectively defined by $j^{\mu,5}=\bar{\psi}\gamma^{\mu}\gamma^5\psi$ and $j^{\mu}=\bar{\psi}\gamma^{\mu}\psi$), which describes a novel interplay between orbital and spin currents. After integrating out the high energy bands we can also obtain different polarizabilities for two bands and this effect can be captured through a term proportional to $\sigma^{0j}E_j$, where $\sigma^{0j}=\{\gamma^0, \gamma^j \}/2i$, and this term also breaks chiral symmetry. If we consider the simple problem of massless fermions and turn off all other chiral symmetry breaking bilinears (e.g., $g_-=0$), we obtain
\begin{eqnarray}
\partial_{\mu }j^{\mu,5}&=&\frac{\epsilon^{\alpha \beta \rho \lambda}}{16 \pi^2} \left (e^2 \mathcal{F}^{V}_{\alpha \beta} \mathcal{F}^{V}_{\rho \lambda} + g_{+}^{2} \mathcal{F}^{A}_{\alpha \beta} \mathcal{F}^{A}_{\rho \lambda}\right), \label{eq:3}\\
\partial_{\mu }j^{\mu}&=&\frac{\epsilon^{\alpha \beta \rho \lambda}}{8 \pi^2} \: eg_{+} \: \mathcal{F}^{V}_{\alpha \beta} \mathcal{F}^{A}_{\rho \lambda} , \label{eq:4}
\end{eqnarray}
where $\mathcal{F}^{V}_{\alpha \beta}$ and $\mathcal{F}^{A}_{\alpha \beta}$ respectively correspond to the flux strength tensors of electromagnetic and axial gauge field. Notice that the non-conservation of electromagnetic current is captured by a \emph{mixed anomaly} term in Eq.~\ref{eq:4}, and rather the modified electromagnetic current $\tilde{j}^{\alpha}=j^{\alpha}-\frac{\epsilon^{\alpha \beta \rho \lambda}}{4 \pi^2}\mathcal{A}_{\beta}\mathcal{F}^{V}_{\rho \lambda}$ will be conserved. For a general chiral gauge field the modification corresponds to \emph{non-dissipative chiral magnetic current and anomalous Hall current}\cite{Goswami,QiWeyl}. Since our chiral vector potential is proportional to the physical magnetic field, $\mathcal{A}_0=0$ (temporal gauge condition). A ``chiral electric field" will appear only for a time dependent physical magnetic field ($\mathbf{B}$), and a spatially varying $\mathbf{B}$ can only induce a ``chiral magnetic field". We also note that a constant magnetic field will correspond to a constant chiral gauge potential, which corresponds to momentum space separation of left and right handed fermions in a Weyl semi-metal \cite{Goswami,QiWeyl}. In the following sections we will show that a mixed anomaly can occur even in the superconducting phase, but its effect can only be observed in the mixed phase. 

\emph{Pairing symmetries :}
 In the presence of superconductivity, we choose a Nambu spinor $\Psi^T=(\psi^{\dagger}, \psi^{T}\gamma^5 \mathcal{C})$, and the $8 \times 8$ pairing Hamiltonian operator can be compactly written as
\begin{equation}
\hat{H}=\left(\begin{array}{c c}
 \hat{h}_{N,1}-\mu \mathbb{1} & \hat{\Delta}\\
\hat{\Delta}^{\dagger} & \hat{h}_{N,2}+\mu \mathbb{1} \end{array} \right)
\label{eq:5}
\end{equation}
where $\hat{h}_{N,1}=\gamma^0(-i\gamma^j\mathcal{D}_j+m+eA_0\gamma^0)$, $\hat{h}_{N,2}=\gamma^0(i\gamma^j\mathcal{D}_{j}^{\ast}-m -eA_0\gamma^0)$ the covariant derivative $\mathcal{D}_j=(\partial_j+ie\mathcal{A}_j)$ and $\mathcal{A}_j$ is the electromagnetic vector potential. The pairing matrix $\hat{\Delta}$ can be decomposed in terms of sixteen $4 \times 4$ matrices, which complete the Clifford algebra \cite{Ohsaku}. The explicit form of the generalized pairing matrix is described by
\begin{equation}
\hat{\Delta}= \{ \Delta^s \mathbb{1} + \Delta^v_\mu \gamma^\mu + \Delta^p  \gamma^5 +\Delta^{a}_{\mu}\gamma^{\mu}\gamma^5 +\Delta_{\mu \nu}^{t}\sigma^{\mu \nu} \} ,
\label{eq:6}
\end{equation}
Under the Lorentz transformations (LT), the pairing bilinears proportional to $\Delta^s$, $\Delta^p$, $\Delta^v_\mu$, $\Delta^{a}_{\mu}$, $\Delta_{\mu \nu}^{t}$ respectively transform as scalar, pseudoscalar, vector, axial vector and antisymmetric tensor. For local pairing (intra-unit cell and momentum independent), the Pauli exclusion principle only allows $\Delta^s$, $\Delta^p$ and $\Delta^v_\mu$. The transformation of the allowed pairings under LT, discrete symmetry operations, chiral transformation, and corresponding quasiparticle spectra (for uniform paired states) are shown in TABLE I.
\begin{widetext}
\onecolumngrid
\begin{table}[htdp]

\caption{Transformation properties of the allowed local pairing channels under the discrete symmetry operations $\mathcal{P}$, $\mathcal{T}$, chiral transformation $\mathcal{U}_{c}$, and the corresponding quasi-particle spectrum within the mean-field approximation. The distinct branches of quasi-particle spectrum are denoted by $\alpha=\pm 1$, and each branch is two-fold degenerate. In Ref.~\onlinecite{FuBerg} the symmetries of possible local pairings for $\mathrm{Cu}_x\mathrm{Bi}_2\mathrm{Se}_3$, have been classified according to the representations of point group $D_{3d}$. The relation between the decomposition in Our $(\Delta^s \mathbb{1}, \Delta^v_0 \gamma^0)$, $\Delta^p i \gamma^5$, $\Delta^v_3\gamma^3$, $(\Delta^v_1\gamma^1,\Delta^v_2\gamma^2)$ respectively correspond to $A_{1g}$, $A_{1u}$, $A_{2u}$ and $E_u$ representations of $D_{3d}$.}

\begin{center}

\begin{tabular}{|l|l|l|l|l|l|l|}

\hline

Pairing & LT & $D_{3d}$ & $\mathcal{T}$ & $\mathcal{P}$ & $\mathcal{U}_{c}$ & Spectrum  \\

\hline

\hline
$\Delta^s$ & Scalar & $A_{1g}$ & + & + & $\times$ &$E_\alpha=\pm \left[ \left( \sqrt{\mathbf{k}^2+m^2} - \alpha \mu \right)^{2} + |\Delta^s|^2 \right]^{1/2}$ \\
\hline
$\Delta^p$ & Pseudoscalar & $A_{1u}$ & + & - & $\times$ &$E_\alpha=\pm \left[ \mathbf{k}^2 + m^2+\mu^2 |\Delta^p|^2 - 2 \alpha \sqrt{ (\mathbf{k}^2+m^2) \mu^2 + m^2 |\Delta^p|^2}  \right]^{1/2}$\\
\hline
$\Delta^{v}_{0}$ & 0-th component of vector & $A_{1g}$ & + & + & \checkmark &$E_\alpha=\pm \left[ \mathbf{k}^2+m^2+\mu^2 + |\Delta^{v}_{0}|^2 - 2 \alpha \sqrt{(\mathbf{k}^2+m^2) \mu^2 + |\Delta^{v}_{0}|^2 \mathbf{k}^2}     \right]^{1/2}$\\
\hline
$\Delta^{v}_{3}$ & 3rd component of vector & $A_{2u}$ & + & - & \checkmark &$E_\alpha=\pm \left[ \mathbf{k}^2 +m^2 + \mu^2 + | \Delta^{v}_{3} |^2 - 2 \alpha
\sqrt{(\mathbf{k}^2+m^2)\mu^2 + m^2 |\Delta^{v}_{3}|^2 + |k_3 \Delta^{v}_{3}|^2 } \; \right]^{1/2}$ \\
\hline
$\Delta^{v}_{1}$ & 1st component of vector & $E_{u}$ & + &- & \checkmark &$E_\alpha=\pm \left[ \mathbf{k}^2 +m^2 + \mu^2 + | \Delta^{v}_{1} |^2 - 2 \alpha
\sqrt{(\mathbf{k}^2+m^2)\mu^2 + m^2 |\Delta^{v}_{1}|^2 + |k_1 \Delta^{v}_{1}|^2 } \; \right]^{1/2}$ \\
\hline
$\Delta^{v}_{2}$ & 2nd component of vector & $E_{u}$ & +  &- & \checkmark &$E_\alpha=\pm \left[ \mathbf{k}^2 +m^2 + \mu^2 + | \Delta^{v}_{2} |^2 - 2 \alpha
\sqrt{(\mathbf{k}^2+m^2)\mu^2 + m^2 |\Delta^{v}_{2}|^2 + |k_2 \Delta^{v}_{2}|^2 } \; \right]^{1/2}$ \\
\hline\hline
\end{tabular}
\end{center}
\end{table}
\end{widetext}
We also display the corresponding classification according to various representations of $D_{3d}$ point group, which was established in Ref.~\onlinecite{FuBerg}.

Notice that only topologically trivial pairing $\Delta^s$ is fully gapped for arbitrary values of the band parameters. In contrast the topological odd parity pairing $\Delta^p$, goes through a phase transition at $\mu^2+\Delta^2 = m^2$ (via a Dirac point) to a topologically trivial state. Only when $\mu^2+\Delta^2 > m^2$ we have a nontrivial $Z$ invariant, $N=\mathrm{sgn}(\Delta^p)$ \cite{ZeroNishida}. Consequently the Majorana surface states can only be found, when this condition is satisfied. For $\mathrm{Cu}_x\mathrm{Bi}_2\mathrm{Se}_3$ and $\mathrm{Sn}_{1-x}\mathrm{In}_x\mathrm{Te}$, the chemical potential lies in the conduction band, this condition is always satisfied. The vector pairing $\Delta_{\mu}^{v}$ are not fully gapped. The space like vectors $\mu=1,2,3$ can have point nodes when $\mu^2+\Delta^2 > m^2$. For example if we consider $j=3$, the nodes will be at $k_z=\pm \sqrt{\mu^2+\Delta^2 - m^2}$. The time like vector pairing exhibits gapless excitations on a three dimensional surface in the momentum space. In the following sections we will only focus on the fully gapped $\Delta^s$ and $\Delta^p$.

\emph{Effective action via chiral rotation:}
The derivation of the effective action in the presence of different chiral symmetry breaking perturbations in the particle-hole channel is a challenging task. However we can gain valuable insight by first considering the pairing of massless Dirac fermions in the absence of any chiral symmetry breaking perturbation in the particle-hole channel. In the chiral basis of the four component spinor $\psi^{T}=(c^{R}_{\uparrow}, c^{R}_{\downarrow}, c^{L}_{\uparrow},c^{L}_{\downarrow})$, the massless Dirac Hamiltonian is block diagonal. The annihilation operators corresponding to the right (R) and the left (L) chiralities are respectively the symmetric and the antisymmetric combinations of the annihilation operators of the even and the odd parity bands $c^{R/L}_{s}=(c^{+}_{s}\pm c^{-}_{s})/\sqrt{2}$. In this basis the trivial and topological pairings are respectively described by $\Delta^s(R^{\dagger}i\sigma_2R^{\ast}+ L^{\dagger}i\sigma_2L^{\ast})$ and $\Delta^p(R^{\dagger}i\sigma_2R^{\ast}+ L^{\dagger}i\sigma_2L^{\ast})$, where we have defined the two-component spinors $R^{T}=(c^{R}_{\uparrow}, c^{R}_{\downarrow})$ and $L^T=(c^{L}_{\uparrow},c^{L}_{\downarrow})$. Therefore $\Delta^{s/p}$ do not not mix $R$ and $L$ sectors and we can consider the pairing problem separately in $R$ and $L$ subspaces. The action in the fermion sector can be written as $S=S_R+S_L$, the action in each subspace takes the following compact form,
\begin{eqnarray}
S_{a}=\int {d^4 x} \: \bar{\Psi}_{a}\left[i\tilde{\gamma}^{\mu}_{a}\left(\partial_{\mu}+A_{\mu,a} \tilde{\gamma}^{5}_{a}\right)-\Delta e^{i\theta_{a}\tilde{\gamma}^{5}_{a}}\right]\Psi_{a}
\label{eq:7}
\end{eqnarray}
where $a=R/L$ denote the chiral subspaces. For the chiral subspaces we have defined two four component Nambu spinors $\Psi_{R}=(R, i\sigma_2 R^{\ast})^{T}$ and $\Psi_{L}=(L, i\sigma_2 L^{\ast})^T$, and two sets of $4\times 4$ gamma matrices $\tilde{\gamma}^{\mu}_{a}$. The explicit forms of these gamma matrices are given by $\tilde{\gamma}^{0}_{R}=\tilde{\gamma}^{0}_{L}=\eta^1 \otimes \sigma^0$, $\tilde{\gamma}^{j}_{R}=-\tilde{\gamma}^{j}_{L}=-i\eta_2 \otimes \sigma^j$, $\tilde{\gamma}^{5}_{R}=\tilde{\gamma}^{5}_{L}=\eta^3 \otimes \sigma^0$. The right and left sectors now have separate gauge fields $A_{\mu, R/L}=e \mathcal{V}_{\mu} \pm g_+ \mathcal{A}_{\nu}$. Also notice that we have allowed two distinct phase angles $\theta_{R/L}$.

In this decoupled chiral Nambu basis, the pairing term appears as a mixture of scalar and pseudoscalar Dirac masses, albeit in the presence of a new chiral gauge field in the Nambu space. This stems from the fact that $\sigma_2 R^{*}$ and $\sigma_2 L^{*}$ respectively transform as left and right chiral spinors. Therefore in each chiral Nambu subspace, the phase of the superconductor appears as the axion angle and the gauge field appears as an axial gauge field. For trivial pairing $\theta_R=\theta_L=\theta$ is the phase corresponding to electromagnetic $U(1)$ symmetry. For the topological pairing $\theta_R=\theta, \ \theta_L= \theta \pm \pi$, where $\theta$ again corresponds to the electromagnetic $U(1)$ phase.

For deriving an effective action of superconducting order parameter by integrating out the fermions, it is a common practice to first absorb the half of the superconducting phase onto the fermion operators, which modifies the gauge field $A_{\mu} \to A_{\mu}-\partial_{\mu}\theta/2 $. In our problem such a phase transformation $\Psi_{a} \to \exp \left( -i/2 \: \theta_{a} \: \tilde{\gamma}_{a}^{5}\right)\Psi_{a}$ constitutes a chiral transformation and a modification of the chiral gauge field. \emph{In this process we should encounter field theory anomalies, which leads to the topological terms in the effective action}. Following Fujikawa we perform the chiral rotation at infinitesimal steps $\Psi_{a} \to \exp \left( -i/2 \: ds \: \theta_{a} \: \tilde{\gamma}_{a}^{5}\right)\Psi_{a}$, and integrate over $s$ from 0 to 1. The topological contribution arises through the Jacobians of the chiral transformation, and are given by
\begin{equation}
J_{a}=\exp \bigg [i \theta_{a} \: \int_0^{1} ds \:  \lim_{M \to \infty} Tr \bigg \{\tilde{\gamma}^{5}_{a}\exp \left(-\frac{D_{a,s}^{2}}{M^2}\right)\bigg \}\bigg],
\label{eq:8}
\end{equation}
where the expression for the Dirac kernel is
\begin{equation}
D_{a}=i\tilde{\gamma}^{\mu}_{a}\left(\partial_{\mu}+\left \{A_{\mu,a}-\frac{\partial_{\mu}\theta_{a}}{2}\right \} \tilde{\gamma}^{5}_{a}\right)-\Delta e^{i (1-ds)\theta_{a}\tilde{\gamma}^{5}_{a}}\label{eq:9}
\end{equation}
The topological terms in this procedure is proportional to $Tr[\tilde{\gamma}^{5}_{a}\sigma^{\mu \nu}_{a} \sigma^{\rho \lambda}_{a}]\mathcal{F}_{\mu \nu, a}\mathcal{F}_{\rho \lambda, a}$. Now notice that
\begin{equation}
\tilde{\gamma}^{5}_{R}=i\tilde{\gamma}^{0}_{R}\tilde{\gamma}^{1}_{R}\tilde{\gamma}^{2}_{R}\tilde{\gamma}^{3}_{R}, \quad \tilde{\gamma}^{5}_{L}=-i\tilde{\gamma}^{0}_{L}\tilde{\gamma}^{1}_{L}\tilde{\gamma}^{2}_{L}\tilde{\gamma}^{3}_{L}\label{eq:10}
\end{equation}
and this difference leads to opposite signs for the topological terms obtained from the right and the left sectors. After performing a standard momentum integral and combining the contributions from both sectors we obtain
\begin{eqnarray}
 &&S_{top}=-\frac{\epsilon^{\mu \nu \rho \lambda}}{64 \pi^2} \int {d^4 x}\left( \theta_R \mathcal{F}_{\mu \nu, R} \mathcal{F}_{\rho \lambda, R}-\theta_L \mathcal{F}_{\mu \nu, L} \mathcal{F}_{\rho \lambda, L}\right)\nonumber  \\ &=& -\frac{\epsilon^{\mu \nu \rho \lambda}}{64 \pi^2}\int {d^4 x}(\theta_R-\theta_L)\left (e^2\mathcal{F}^{V}_{\mu \nu} \mathcal{F}^{V}_{\rho \lambda} + g_{+}^{2} \mathcal{F}^{A}_{\mu \nu} \mathcal{F}^{A}_{\rho \lambda}\right)\nonumber \\
&&-\frac{\epsilon^{\mu \nu \rho \lambda}}{32 \pi^2}\: eg_{+}\int {d^4 x}(\theta_R+\theta_L)\: \mathcal{F}^{V}_{\mu \nu} \mathcal{F}^{A}_{\rho \lambda} \nonumber \\ \label{eq:11}
\end{eqnarray}
which is the main result of this paper.

We first consider topologically trivial superconductor, and using $\theta_R=\theta_L=\theta$ in Eq.~\ref{eq:11} we obtain,
\begin{eqnarray}
S_{top,s}=-\frac{\epsilon^{\mu \nu \rho \lambda}}{16 \pi^2} \: e g_{+} \:\int {d^4 x} \: \theta \: \mathcal{F}^{V}_{\mu \nu} \mathcal{F}^{A}_{\rho \lambda} \label{eq:12}
\end{eqnarray}
On the other hand for topological odd parity superconductor, we substitute $\theta_R-\theta_L=\pi$ and find
\begin{widetext}
\begin{eqnarray}
S_{top,p}=\int {d^4 x} \bigg[-\frac{\epsilon^{\mu \nu \rho \lambda}}{16 \pi^2} \: e g_{+} \: \theta \: \mathcal{F}^{V}_{\mu \nu} \mathcal{F}^{A}_{\rho \lambda} -\frac{\epsilon^{\mu \nu \rho \lambda} \pi }{64 \pi^2}\bigg (e^2\mathcal{F}^{V}_{\mu \nu} \mathcal{F}^{V}_{\rho \lambda} + g_{+}^{2} \mathcal{F}^{A}_{\mu \nu} \mathcal{F}^{A}_{\rho \lambda}+2eg_{+}\mathcal{F}^{V}_{\mu \nu}\mathcal{F}^{A}_{\rho \lambda}\bigg)\bigg]\nonumber \\ \label{eq:13}
\end{eqnarray}
\end{widetext}
Notice that for massless Dirac fermions, both superconductors have a mixed anomaly present in the bulk. The axion angle $\theta$ for the mixed anomaly is a dynamic phase variable, and will modify the Maxwell's equations in the bulk. This represents the interplay between charge and spin currents. In addition to the bulk term, the odd parity pairing has an addition anomaly term with constant axion angle $\pi$, which is inoperative inside the bulk. This term only becomes important at the surface, and corresponds to the existence of surface Majorana fermions. The BdG nature of the quasiparticles has contributed by a multiplicative factor of 1/2 to the conventional formula of anomaly. Also notice that mixed anomaly term does contribute to the surface action. If we do not consider the Zeeman term (chiral gauge field), there will be no bulk anomaly, and only surface anomaly term for the electromagnetic gauge field will be present for odd parity pairing. In that case our Eq.~\ref{eq:13} will simplify to
\begin{equation}
S_{top,p}=-\frac{e^2 \pi \epsilon^{\mu \nu \rho \lambda}}{64 \pi^2} \int {d^4 x} \mathcal{F}^{V}_{\mu \nu} \mathcal{F}^{V}_{\rho \lambda}\label{eq:14}
\end{equation}
which precisely agrees with the topological term derived in Ref.~\onlinecite{QiWitten}, using (4+1)-dimensional regularization. In the topological insulator where electromagnetic charge is a conserved quantity, the existence of such an axion term in the electrodynamics (with twice the coefficient) corresponds to half-integer quantum Hall effect of the surface Dirac fermions \cite{AxionHughes}. However in the superconductor electromagnetic gauge symmetry is broken, and electromagnetic charge is not a conserved quantity. Therefore, there will be no surface quantum Hall effect in the present problem.

The conventional terms of the Landau-Ginzburg theory, describing the Meissner effects etc. is obtained by integrating out the transformed fermions, assuming slowly varying chiral supercurrents $(2A_{\mu,\alpha}-\partial_{\mu}\theta_{\alpha})$ and $\Delta(x)$. In the London limit we can neglect the fluctuations of the amplitude $\Delta(x)$, and obtain
\begin{eqnarray}
&& S_{GL}=\int {d^4 x} \bigg \{\frac{\rho}{2}\bigg[\left(\partial_{\mu}\theta-2e\mathcal{V}_{\mu}\right)^2+4g_{+}^{2}\mathcal{A}_{\mu}^{2}\bigg]+\frac{1}{4}\bigg((\mathcal{F}^{V}_{\mu \nu})^2\nonumber \\&& \hspace{5cm}+(\mathcal{F}^{A}_{\mu \nu})^2\bigg)\bigg \}\label{eq:15},
\end{eqnarray}
In the above equation we have also ignored the mismatch between Fermi velocity and speed of light. Now combining $S_{GL}$ with $S_{top}$ we obtain the total effective action for the superconducting order parameter in the London limit. Now taking a derivative of the total action in the bulk with respect to $\mathcal{V}_{\mu}$, we find that the conventional supercurrent $2 e\rho(\partial_{\mu}\theta-2e\mathcal{V}_{\mu})$ is not conserved, when we are in the vortex/mixed phase. It turns out that topological superconductor has fermion zero modes in the vortex core, which provides an additional path for transport, and the mixed anomaly is precisely tied to this effect. Motivated by this we now proceed to explicit computation of fermion zero modes in the dilute vortex limit. Fermion zero modes in the absence of Zeeman coupling have been recently addressed by various authors. However we will account for the general Zeeman coupling and its possible variation due to Meissner screening away from the vortex core in the following section.

\emph{Vortex zero modes of generic Dirac Hamiltonian with odd parity pairing :}
We consider a static line vortex along the $z$ direction, and solve the zero mode problem for the planar part. In the absence of Dirac mass, chemical potential, and the Zeeman couplings the planar problem corresponds to two copies of Jackiw-Rossi Hamiltonians, which belongs to class $BDI$ and we obtain $2n$ number of Majorana zero modes, where $n$ is the vorticity \cite{ZeroJackiw}. In the presence of the additional perturbations mentioned above, the problem belongs to class $D$ and only for odd vorticity we can find $2$ Majorana zero modes \cite{TeoKane}, which is tied to a $Z_2$ index theorem \cite{Tewari,Fukui}.

Let us consider a particular profile of the electromagnetic gauge potential ${\cal V}_\phi= \frac{1}{2 \lambda^2} r$ if $r<\lambda$, defining the core of the vortex and ${\cal V}_\phi =1/2r$ outside ($r > \lambda$), if one wishes to commit to the spherically symmetric gauge. Then the magnetic field ($B$) is finite and constant, $B=\frac{1}{2 \lambda^2}$ only inside the vortex core, while it vanishes outside. The superconducting order parameter vanishes smoothly inside the core as $r \rightarrow 0$, and satutares at $\Delta_0$ as $r \rightarrow \infty$, otherwise arbitrary. Upon including the Dirac mass ($m_k$), Zeeman couplings ($h_1, h_2$) one set of coupled differential equations of zero energy modes in the plane perpendicular to the applied field read as

\begin{eqnarray}\label{majorana1}
 e^{-i \phi} \left( \partial_z+{\cal V}_\phi \right) \Lambda^{-}_\downarrow + i a (r) \; \Lambda^{+}_\uparrow &+& i \Delta_r e^{-i \phi} \left( \Lambda^{-}_\downarrow\right)^\dagger = 0, \nonumber \\
i e^{i \phi} \left( \partial_{\bar{z}}-{\cal V}_\phi \right)\Lambda^{+}_\uparrow + b (r) \; \Lambda^{-}_\downarrow &+& \Delta_r e^{-i \phi} \left( \Lambda^{+}_\uparrow\right)^\dagger =0,
\end{eqnarray}
at finite chemical potential ($\mu$), where $\partial_z=\partial_r - \frac{i}{r} \partial_\phi$, $\partial_{\bar{z}}=\partial^*_z$ and $a(r)=m_k+\mu+h_1(r)$, $b(r)=m_k+h_2(r)-\mu$, and in what follows we set $m_k=m$ (constant). In terms of g-factors, which can in principle be different in two bands, one can write $h_i (r)= g_i B(r)$, where $i=1,2$. Therefore, the Zeeman coupling is only finite inside the vortex core, and vanishes outside. The other set of coupled differential equations for the zero mode is

\begin{eqnarray}\label{majorana2}
 e^{-i \phi} \left( \partial_z+{\cal V}_\phi \right) \Lambda^{+}_\downarrow - i d(r) \; \Lambda^{-}_\uparrow &+& i \Delta_r e^{-i \phi} \left( \Lambda^{+}_\downarrow\right)^\dagger=0, \nonumber \\
(i) e^{i \phi} \left( \partial_{\bar{z}}-{\cal V}_\phi \right)\Lambda^{-}_\uparrow - c(r) \; \Lambda^{+}_\downarrow &+& \Delta_r e^{-i \phi} \left( \Lambda^{-}_\uparrow\right)^\dagger=0,
\end{eqnarray}
where $c(r)=m_k-h_1 (r)+\mu,\; d(r)=m_k-h_2 (r)-\mu$. The remaining \emph{four} equations are related to these by Hermitian conjugation.

Two BdG-Majorana modes in Nambu-Dirac basis, defined as $\Psi = \left[ \Psi^\top_p (+ \vec{k}),\Psi^\top_h (-\vec{k}) \right]$, where $\Psi^\top_p(+\vec{k})=\Psi^\top(\vec{k})$ and $\Psi^\top_h(-\vec{k})=\Psi_p(\vec{k})$, otherwise
\begin{equation} \label{4compspinor}
\Psi^\top(\vec{x})= \left[ |\Lambda^+,+\frac{1}{2} \rangle, |\Lambda^-,+\frac{1}{2} \rangle, |\Lambda^+,-\frac{1}{2} \rangle, |\Lambda^-,-\frac{1}{2} \rangle \right](\vec{x}),
\end{equation}
take the form
\begin{equation}
| \Psi^0_1 \rangle ={\cal R}(r) \left(
\begin{array}{c}
g(r) e^{-i \phi}\\ 0 \\ 0 \\ i f(r) \\ i g(r) e^{i \phi}\\ 0 \\ 0 \\ f(r)
\end{array} \right); | \Psi^0_2 \rangle ={\cal R}(r) \left( \begin{array}{c}
0\\ i q(r)\\ p(r) e^{-i \phi}\\ 0\\ 0\\ q(r)\\ i p(r) e^{i \phi}\\ 0 \end{array} \right),
\end{equation}

where ${\cal R}(r)= \exp{\left(-i \frac{\pi}{4} -\int^r_0 \Delta_{r'} dr' \right)}$. For $r/2 \lambda \ll 1$
\begin{equation}\label{majorana1-insidecore}
g(r)=c_1 \; I_1 \bigg[ r \sqrt{a b}\bigg], \quad f(r)= c_1 \; \sqrt{\frac{a}{b}} \; I_0 \bigg[ r \sqrt{a b} \bigg],
\end{equation}
 where $I_j$, with $j=0,1$ are the modified Bessel functions of order $j$. On the other hand, deep within the vortex core
\begin{equation}\label{majorana2-insidecore}
q(r)=\tilde{c}_1 \; I_1 \bigg[ r \sqrt{c d}\bigg], \quad p(r)= \tilde{c}_1 \; \sqrt{\frac{d}{c}} \; I_0 \bigg[ r \sqrt{c d} \bigg].
\end{equation}
Outside the core of the vortex ($r>\lambda$)
\begin{eqnarray}\label{majorana1-outsidecore}
g(r)&=&c_2 I_{1/2} \left(r \sqrt{\beta \eta} \right) + c_3 I_{-1/2} \left( r \sqrt{\beta \eta} \right), \nonumber \\
 f(r)&=& \sqrt{\frac{\beta}{\eta}} \left[ c_2 I_{-1/2} \left(r \sqrt{\beta \eta} \right) + c_3 I_{1/2} \left( r \sqrt{\beta \eta} \right) \right].
\end{eqnarray}
where $\beta=m +\mu$, and $\eta=|m-\mu|$. In the regime, $g(r) \equiv q(r)$ and $f(r) \equiv p(r)$. Two out of three arbitrary constants, for each of the BdG-Majorana solutions can be fixed by equating the value of the functions, and their first derivatives at $r=\lambda$. The remaining constant can be determined from the normalization condition. The modified Bessel functions $I_{\pm 1/2}(\alpha r) \propto e^{\alpha r}/r\alpha$, grow exponential at large distances \cite{ZeroHerbut,ZeroTagliacozzo}. Therefore, for normalizable zero energy modes can only be found if the condition
\begin{equation}
\mu^2 + \Delta^2_0 \geq m^2,
\end{equation}
is satisfied. It is worth noticing that the nomalizability of the solutions is independent of the Zeeman couplings.

We now account for the $\partial_3$ and $A_3$ along the vortex line perturbatively. This perturbation reads as ${\cal H}_3= \left( \tau_0 \left( -i \partial_3\right) + \tau_3 A_3 \right)\otimes \gamma^0 \gamma^3$ and keeps the BdG-Majorana sub-space invariant. In this basis ${\cal H}_3$ is purely off diagonal, and takes the form
\begin{equation}
 {\cal H}_3= -v_1 \sigma_2 (i) \left( \partial_3 -i A_3 \right)- v_2 \sigma_1 (i) \left( \partial_3 - \frac{v_1}{v_2} i A_3 \right).\label{1dh}
 \end{equation}

Two parameters $(v_1,v_2)$ of the one dimensional \emph{helical} BdG-Majorana modes depend on the exact profile of the vortex, and other microscopic parameters (Dirac mass, chemical potential etc.). However, in the absence of any Zeeman coupling, i.e. $h_1=h_2=0$, one finds $v_1=v_2$. The above Hamiltonain in Eq.~\ref{1dh} can be diagonalized in the basis of symmetric and anti-symmetric combinations of $|\Psi^0_1 \rangle$ and $| \Psi^0_2 \rangle$, yielding
\begin{equation}
 {\cal H}_3 \rightarrow - v_1 i \sigma_3 \left( \partial_3 -i A_3 \right).
 \end{equation}
Notice that gaguge field is coupling in conventional way (unlike Nambu space), which reflects the quasiparticle inside the core describes normal state. Since we have both left and right movers (in one dimension) coupled to the electromagnetic field, we do not have any electromagnetic anomaly in one dimension. This justifies that there is no builk anomaly in the absence of Zeeman/chiral gauge field. In the presence of the Zeeman coupling we find the both electromagnetic and chiral gauge field couplings for the one dimensional modes, which leads to the one dimensional version of the chiral anomaly. The presence of the bulk anomaly term is crucial to feed the current to one dimensional world via Callan-Harvey mechanism \cite{CallanHarvey}.

In the absence of the chiral symmetry breaking perturbations (Dirac mass, $g_-$) in the particle hole channel, fermion zero modes are also present for the topologically trivial pairing $\Delta^s$. However these zero modes get removed by such chiral symmetry breaking perturbations \cite{Nishida}. Therefore we believe the mixed anomaly term will be absent for trivial superconductor under generic conditions. In contrast the zero modes of the topological pairing are robust under generic conditions, and this suggests that mixed anomaly term will be present for topological pairing, even in the presence of generic perturbations. However the coefficient instead of being $e g_{+}$ will have more involved model dependence. In contrast a Dirac mass in the absence of time reversal symmetry breaking perturbations, will modify the surface term by a step function $\Theta(\mu^2+\Delta^2-m^2)$, which is the condition for finding massless surface Majorana fermions. The presence of time reversal symmetry breaking perturbations will induce a gap for the surface Majorana fermions.

\emph{Experimental consequences :} We have found a general effective action for the topological superconducting state, which suggests an intriguing interplay between charge and spin currents. The bulk anomaly terms (proportional to the dynamic phase variable) will have nontrivial effects on the electrodynamics equations in the vortex phase. However these will be appreciable only for time-dependent electromagnetic fields. The derived surface anomaly terms will also play important role in the electrodynamic properties of Josephson junction measurements. The presence of robust fermion zero modes in the dilute vortex limit suggests a \emph{T-linear specific heat} in the mixed state. The anomaly terms in the effective action suggests that such behavior should be present throughout the mixed phase. More detailed derivation of the effective action under generic conditions, and a thorough analysis of the electrodynamics of the mixed state will be presented in a forthcoming publication \cite{Broy}.

P. G. and B. R. were supported at the National High Magnetic Field Laboratory by NSF Cooperative Agreement No.DMR-0654118, the State of Florida, and the U. S. Department of Energy.


\begin{thebibliography}{}

\bibitem{Review1} M. Z. Hassan and C. L. Kane, Rev. Mod. Phys. {\bf 82}, 3045 (2010) and references therein.

\bibitem{Review2} M. Z. Hassan and J. E. Moore, Ann. Rev. Cond. Matt. Phys. {\bf 2}, 55 (2011) and references therein.

\bibitem{Review3} X. L. Qi and  S. C. Zhang, Rev. Mod. Phys. {\bf 83}, 1057 (2011) and references therein.

\bibitem{Volovik} G. E. Volovik, {\it The Universe in a Helium Droplet} (Clarendon Press, 2003).

\bibitem{FuBerg} L. Fu and E. Berg, Phys. Rev. Lett. \textbf{105}, 097001 (2010).

\bibitem{ObWray1} A.L. Wray, \emph{et al.}, Nature Phys. {\bf 6}, 855 (2010).

\bibitem{ObKriener} M. Kriener, \emph{et al.}, Phys. Rev. Lett. {\bf 106}, 127004 (2011).

\bibitem{ObWray2} A. L. Wray, \emph{et al.}, Phys. Rev. B {\bf 83}, 224516 (2011).

\bibitem{ObSasaki} S. Sasaki, \emph{et al.}, arxiv:1208.0059

\bibitem{Altland} A. Altland and M. R. Zirnbauer, Phys. Rev. B {\bf 55}, 1142 (1997).

\bibitem{TenfoldRyu}A. P. Schnyder, S. Ryu, A. Furusaki, A. W. W. Ludwig, Phys. Rev. B {\bf 78}, 195125 (2008).

\bibitem{PCSasaki} S. Sasaki, \emph{et al.}, Phys. Rev. Lett. {107}, 217001 (2011).

\bibitem{PCKirzhner}T. Kirzhner, \emph{et al.}, Phys. Rev. B {\bf 86}, 064517 (2012).

\bibitem{PCChen} X. Chen, \emph{et al.}, arxiv:1210.6054

\bibitem{MagDas} P. Das, Y. Suzuki, M. Tachiki, K. Kadowaki, Phys. Rev. B {\bf  83}, 220513 (2011).

\bibitem{MagBay}T. V. Bay, \emph{et al.}, Phys. Rev. Lett. {\bf 108}, 057001 (2012).

\bibitem{MagKriener} M. Kriener, \emph{et. al.}, arxiv:1206.6260

\bibitem{ObLevy} N. Levy, \emph{et al.}, arxiv:1211.0267

\bibitem{ThSHao} L. Hao, T. K. Lee, Phys. Rev. B {\bf 83}, 134516 (2011).

\bibitem{ThSYamakage} A. Yamakage, K. Yada, M. Sato, Y. Tanaka, Phys. Rev. B {\bf  85}, 180509(R) (2012).

\bibitem{ThSMichaeli} K. Michaeli, L. Fu, Phys. Rev. Lett. {\bf  109}, 187003 (2012).

\bibitem{ThSHsieh} T. Hsieh, L. Fu, Phys. Rev. Lett. {\bf 108}, 107005 (2012).

\bibitem{ThSChung} S. B. Chung, J. Horowitz, X. L. Qi, arxiv:1208.3928

\bibitem{GravQi}Z. Wang, X. L. Qi, S. C. Zhang, Phys. Rev. B \textbf{84}, 014527 (2011).

\bibitem{GravRyu1}S. Ryu, J. E. Moore, A. W. W. Ludwig, Phys. Rev. B {\bf 85}, 045104 (2012).

\bibitem{GravRyu2}K. Nomura, S. Ryu, A. Furusaki, N. Nagaosa, Phys. Rev. Lett. {\bf 108}, 026802 (2012).

\bibitem{BFHanson} T. H. Hanson, A. Karlhede, M. Sato, N. Jour. Phys. {\bf 14}, 063017 (2012).

\bibitem{QiWitten} X. L. Qi, E. Witten, S. C Zhang, arxiv:1206.1407 (2012).

\bibitem{Fujikawa} K. Fujikawa, and H. Suzuki, {\it Path integrals and quantum anomalies} (Oxford University Press 2004).

\bibitem{LiuZeeman}C. X. Liu \emph{et al.}, Phys. Rev. B {\bf 82}, 045122 (2010).

\bibitem{Goswami} P. Goswami, and S. Tewari, arXiv:1210.6352 (2012).

\bibitem{QiWeyl}C. X. Liu, P. Ye, X. L. Qi, arXiv:1204.6551 (2012).

\bibitem{Ohsaku} T. Ohsaku, Phys. Rev. B {\bf 65}, 024512 (2002).

\bibitem{ZeroNishida} Y. Nishida, Phys. Rev. D {\bf  81}, 074004 (2010).

\bibitem{AxionHughes} X. L. Qi, T. L. Hughes, and S. C. Zhang, Phys. Rev. B \textbf{78}, 195424 (2008).

\bibitem{ZeroJackiw}R. Jackiw and P. Rossi, Nucl. Phys. B {\bf 190}, 681 (1981).

\bibitem{TeoKane} J. C. Y. Teo and C.L. Kane, Phys. Rev. B \textbf{82}, 115120 (2010).

\bibitem{Tewari} S. Tewari, S. Das Sarma, D. H. Lee, Phys. Rev. Lett. {\bf 99}, 037001 (2007).

\bibitem{Fukui} T. Fukui and T. Fujiwara, Phys. Rev. B {\bf 82}, 184536 (2010).

\bibitem{ZeroHerbut} I.F. Herbut, C-K. Lu, Phys. Rev. B {\bf 82}, 125402 (2010).

\bibitem{ZeroTagliacozzo} A.Tagliacozzo, P. Lucignano, F. Tafuri, Phys. Rev. B {\bf  86}, 045435 (2012).

\bibitem{CallanHarvey}C. G. Callan and J. A. Harvey, Nucl. Phys. B \textbf{250}, 427 (1985).

\bibitem{Broy}P. Goswami and B. Roy (to be published).

\end{thebibliography}
\end{document}